\documentclass{appolb}
\usepackage{graphicx}
\begin{document}
\title{Underwater detection of dangerous substances: status the SABAT project%
\thanks{Presented at 2$^{\mathrm{nd}}$ Jagiellonian Symposium on Fundamental and Applied Subatomic
Physics (Krak{\'o}w, June 3 - 11, 2017). }%
}
\author{M.~Silarski$^{a}$, P.~Sibczy{\'n}ski$^{b}$, Sz.~Nied{\'z}wiecki$^{a}$, S.~Sharma$^{a}$, J.~Raj$^{a}$,
P.~Moskal$^{a}$
\address{$^{a}$Institute of Physics, Jagiellonian University,\\ 
{\L}ojasiewicza 11, 30-348 Krak{\'o}w, Poland}
\address{$^{b}$ National Centre for Nuclear Research, So{\l}tana 7, 05-400 Otwock, Poland}
	}
\maketitle
\begin{abstract}
The Neutron Activation Analysis (NAA) plays an exceptional role in the modern nuclear engineering, especially in
detection of hazardous substances. However, in the aquatic environment, there are still many problems
to be solved for effective usage of this technique. We present status of SABAT
(Stoichiometry Analysis By Activation Techniques), one of the projects aiming
at construction of an underwater device for non-invasive threat detection based on the NAA.
\end{abstract}
  
\section{Introduction}
In twenty-first century the risk of terrorists attack is constantly growing around the world. This situation
put us in a need to develop more effective methods to detect potential threats and
smuggling of illicit materials, e.g. explosives or drugs. New methods are needed particularly for shores and ports
protection and monitoring. They are very important also in view of environmental protection of sea areas of
intensive warfare, e.g. Baltic Sea~\cite{Surveyor,czasmorza}. Over 250 kilotons of munition were sunk into Baltic
Sea, mostly explosives,
but also many chemical agents. It was estimated that if only 16~$\%$ of the sunken ammunition was released into
the Baltic the life in the sea and at its shores would be entirely ruined for the next 100
years~\cite{Silarski2016}. Precise knowledge of location and amount of these hazardous substances
is crucial for a proper planning to deal with such ecological bomb.\\
The presently used methods of detection are based on sonars followed by divers inspection,
which is not effective and rather expensive and puts the divers into danger due to unknown composition of
the suspected items found by the sonar. One of the methods which has a big potential to substitute or support
sonars is the Neutron Activation Analysis (NAA). It is based on fast or thermal neutron beams which excite nuclei
of investigated substance. Detection of characteristic gamma quanta emitted in de-excitation of the nuclei allows
one to identify the stoichiometry of the substance and determine if it is dangerous~\cite{moskalAnn,sabat1}.
There are several designed and produced devices utilizing the NAA for detecting dangerous substances on the
ground~\cite{phoenix,maglich,euritr,sodern,swan}, but in the water one needs to overcome many difficulties
connected to neutron attenuation and huge background from oxygen and hydrogen. This background can be reduced
for example by using neutron generators with associated $\alpha$ particle measurement and by decreasing
the distance between the inspected object and the detector~\cite{Surveyor}. There are also solutions based 
on low energy neutrons which are moderated in water before reaching the tested object. The detector is then
registering gamma quanta originating from thermal neutron capture~\cite{Silarski2016}.\\
In this articles we present status of the design of a NAA-based device within the SABAT project, where we use
guides for the neutron beam and gamma quanta emitted towards the detector. This method not only reduces the
background from water but also may provide detection of dangerous substances hidden deep in the bottom of the
sea and may allow determination of the density distribution of the dangerous substance in the tested object. 
Detailed description of the project can be found in~\cite{ActaA} and~\cite{patent1}.
\section{Preliminary MCNP simulations of the SABAT detection system}
\begin{figure}
\centerline{%
\includegraphics[width=8.cm]{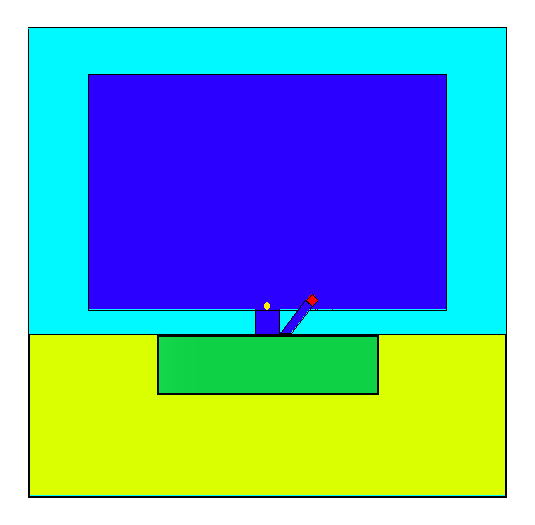}}
\caption{Exemplary scheme of the simulated SABAT sensor geometry. An submarine (blue rectangle) with dimensions
300 x 300 x 200 cm$^3$ contains the neutron source (yellow dot) and gamma quanta detector (red). The neutron and
$\gamma$ quanta guides (also in blue) are simulated as 20 x 20 x 10 cm$^3$ cuboid and 26 cm long polyhedron with 
20 x 7.56 cm$^2$ and 16.7 x 20 cm$^2$ bases, respectively. An container with mustard gas with dimensions
194 x 50 x 50 cm$^3$ (in green) is placed inside the bottom of the sea represented by yellow rectangle
(400 x 400 x 151.5 cm$^3$). Both the submarine and guides are filled with air under normal pressure.}
\label{fig1}
\end{figure}

To design and optimize the SABAT detector in terms of high sensitivity and interrogation time
we have performed Monte Carlo simulations using the general Monte Carlo N-Particle Transport
Code (MCNP) package~\cite{mcnp}. We studied geometry of the system including the relative
positions of gamma quanta detector and the neutron generator, and the optimal guides dimensions. Moreover, one
needs to determine the optimal material composition of the whole device.
The simulated setup is shown in Fig.~\ref{fig1}. An submarine (blue rectangle) is positioned above
the seabed (in yellow) inside which there is an container with hazardous material.
The gamma quanta detector and neutron source (generator) are placed inside the submarine and are connected
to guides filled with air.
Since we optimize the sensor for the detection of war remnants, in particular chemical agents, we assume that the
simulated shell contains mustard gas (C$_4$H$_8$Cl$_2$S). The generator emits neutrons isotropically and
only a small fraction which travels inside the neutron guide reaches the interrogated item without
any interaction. These neutrons may activate the mustard gas via inelastic scattering or neutron capture
which results in emission of the characteristic gamma quanta which are eventually registered by the detector.
It is usually a semiconductor detector, e.g. silicon or HPGe (High Purity Germanium), since it is characterized by very good
energy resolution and detection efficiency. On the other hand, they need a cooling system which limits mobility
of the whole device. Moreover, HPGe detectors are known to be sensitive to neutron flux, which results
in a degradation of the energy resolution over time. As an alternative material one usually uses scintillators.
Among many commercially available scintillator materials we have considered so far BGO, NaI(Tl) and LaBr$_3$(Ce).
BGO is more efficient than the other two materials due to high density, but it has worst energy
resolution~\cite{panda}. The best choice, according to Refs.~\cite{panda,uncoss2},
is the LaBr$_3$(Ce) with a good energy and time resolutions and providing detection efficiency close to
the one of HPGe detectors.
\begin{figure}
%\centerline{%
\includegraphics[width=11.cm]{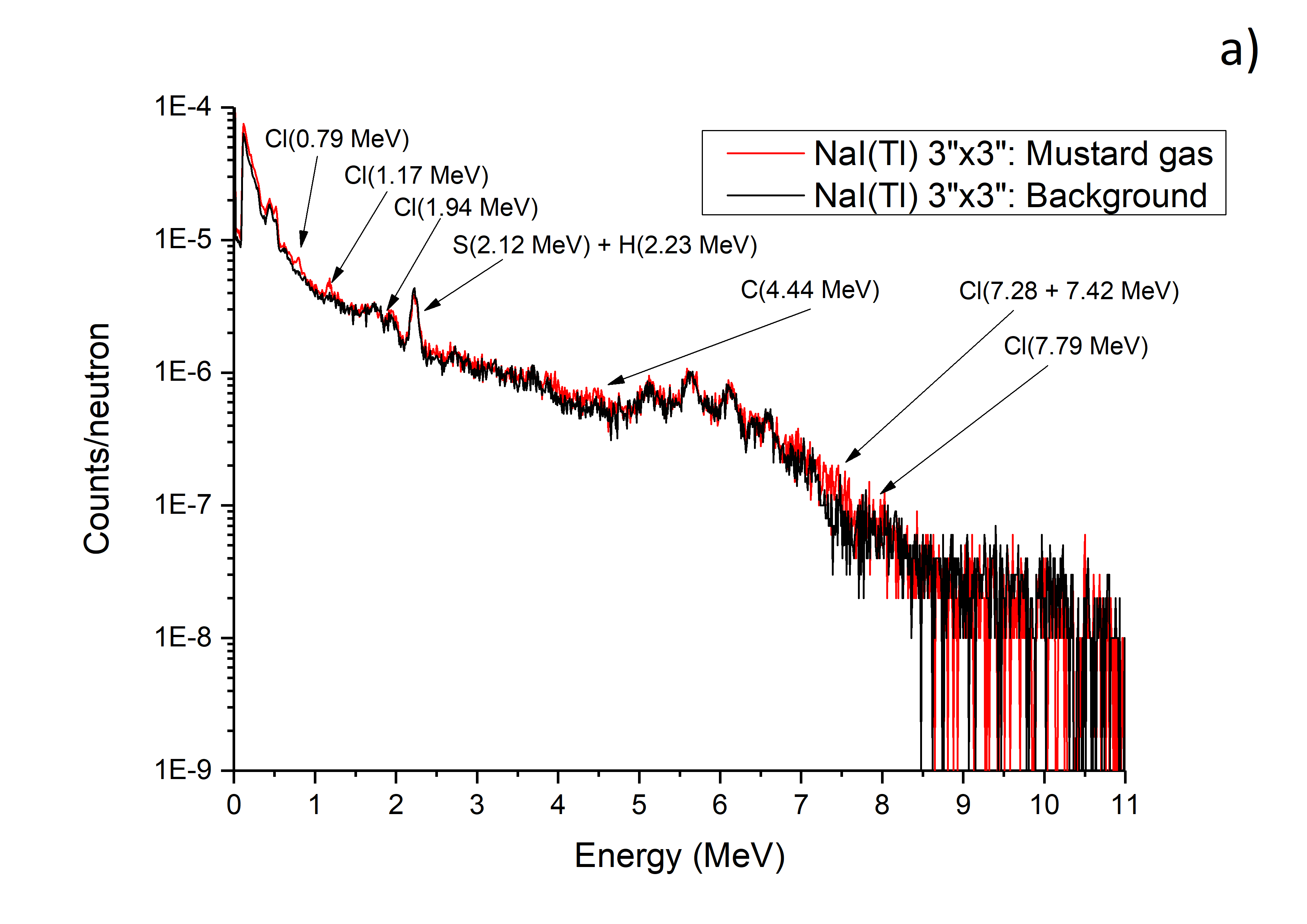}
\includegraphics[width=11.cm]{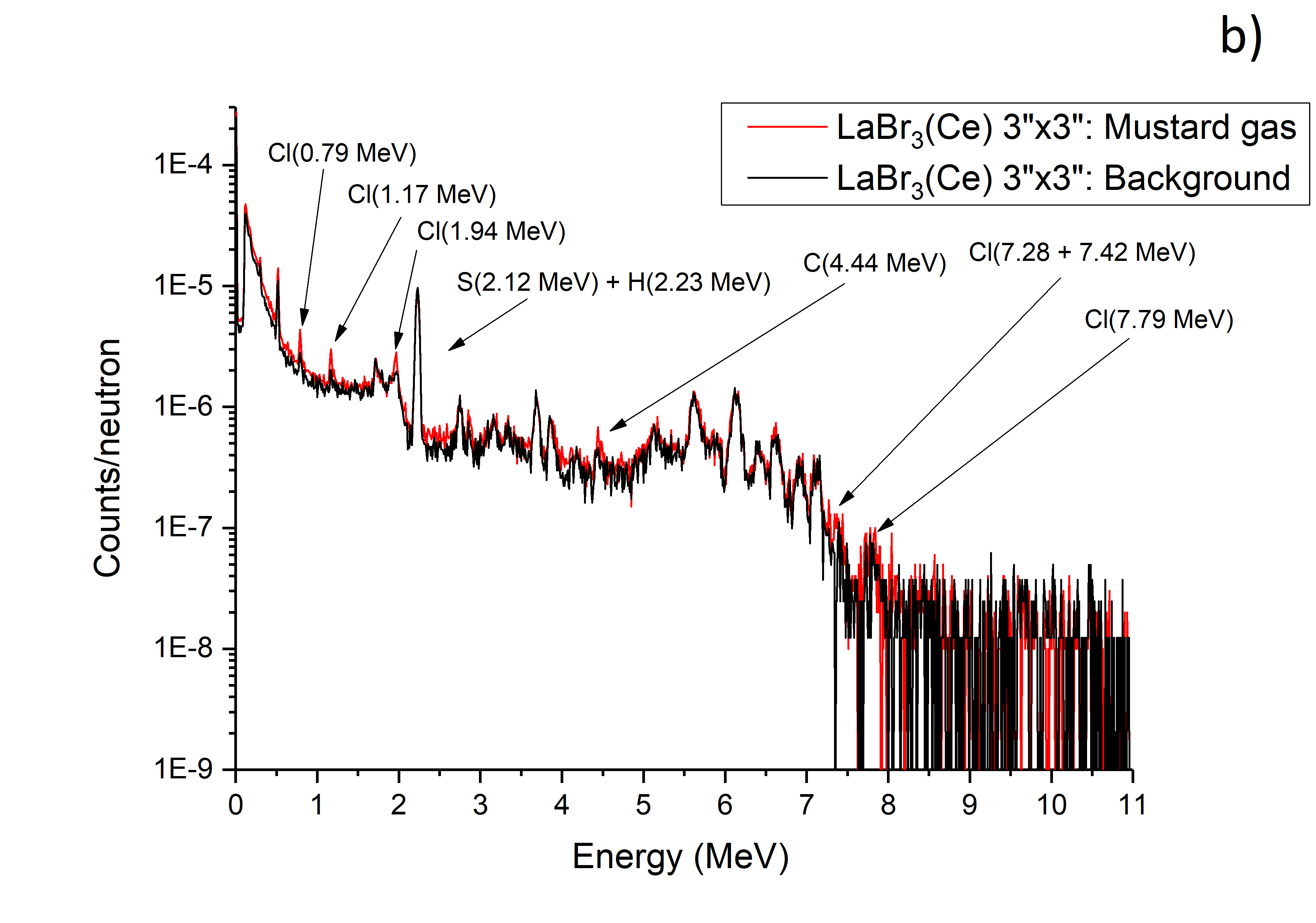}
%}
\caption{Energy distributions simulated assuming the $\gamma$ quanta detector made from a) 3"x3" NaI(Tl)
and b) 3"x3" LaBr$_3$(Ce). The black curve represents simulations with mustard gas container while in red
we present background simulation when the gas is replaced by sand.
}
\label{fig2}
\end{figure}
Exemplary energy spectra of the activated mustard gas obtained with 10$^8$ generated neutrons and assuming
the gamma quanta detector made out of NaI(Tl) and LaBr$_3$(Ce) are shown in Fig.~\ref{fig2} a) and b),
respectively.In the simulations we assumed that the neutron generator works in continuous beam mode.
As expected for both distributions we observe huge background with oxygen peaks around 5-6.5 MeV.
The sulfur line at 2.12~MeV is completely covered by very close hydrogen peak. Similarly, the 4.4~MeV line
of carbon is overwhelmed by the background, mostly by the 4.5~MeV characteristic $\gamma$ quanta
of silicon~\cite{endf,ensdf}.
We see instead small peaks from chlorine at 0.79, 1.16, 1.94 MeV, 7.42, 7.80 and 8.58 MeV.
As one can see the NaI(Tl) detector does not
provide a clear C peak and the Chorine lines are much more smeared. The simulations indicate that the
performance of this detector could be improved by increasing its size but the identification
of potential threats will be anyway much more difficult than in the case of LaBr$_3$(Ce).
To decrease the background from water we introduced a 5~cm thick led shield around the gamma quanta detector,
leaving open only the face connected to the $\gamma$ quanta guide.
%
%Since the detector is relatively close to the neutron generator the shield, apart from suppressing gamma
%lines from water, introduced also additional background from Pb lines, in particular at 7.37~MeV.
%
Unfortunately we did not see any improvement in the sensitivity of the simulated sensor.\\
%Application of additional shielding from neutrons with led or borated polyethylene is also disadvantageous since
%we loose sensitivity to carbon.\\
The performed simulations allowed us also to analyze time structure of the activation processes during
the interrogation of the suspected object. In general there are two gamma quanta groups contributing to
the energy distributions measured by the detector. First group consists of the prompt gammas hitting
the detector within tens of nanoseconds after the neutron emission, depending on the distances between
the activated object, neutron source and the detector. These gamma quanta originate mostly from the nucleus
excitation by the inelastic scattering. The other group of gammas is registered much later and comes from
the thermal neutron capture. The 14.1~MeV
neutrons need to be moderated before they can be captured which takes relatively long time.
\begin{figure}
%\centerline{%
\includegraphics[width=11.0cm]{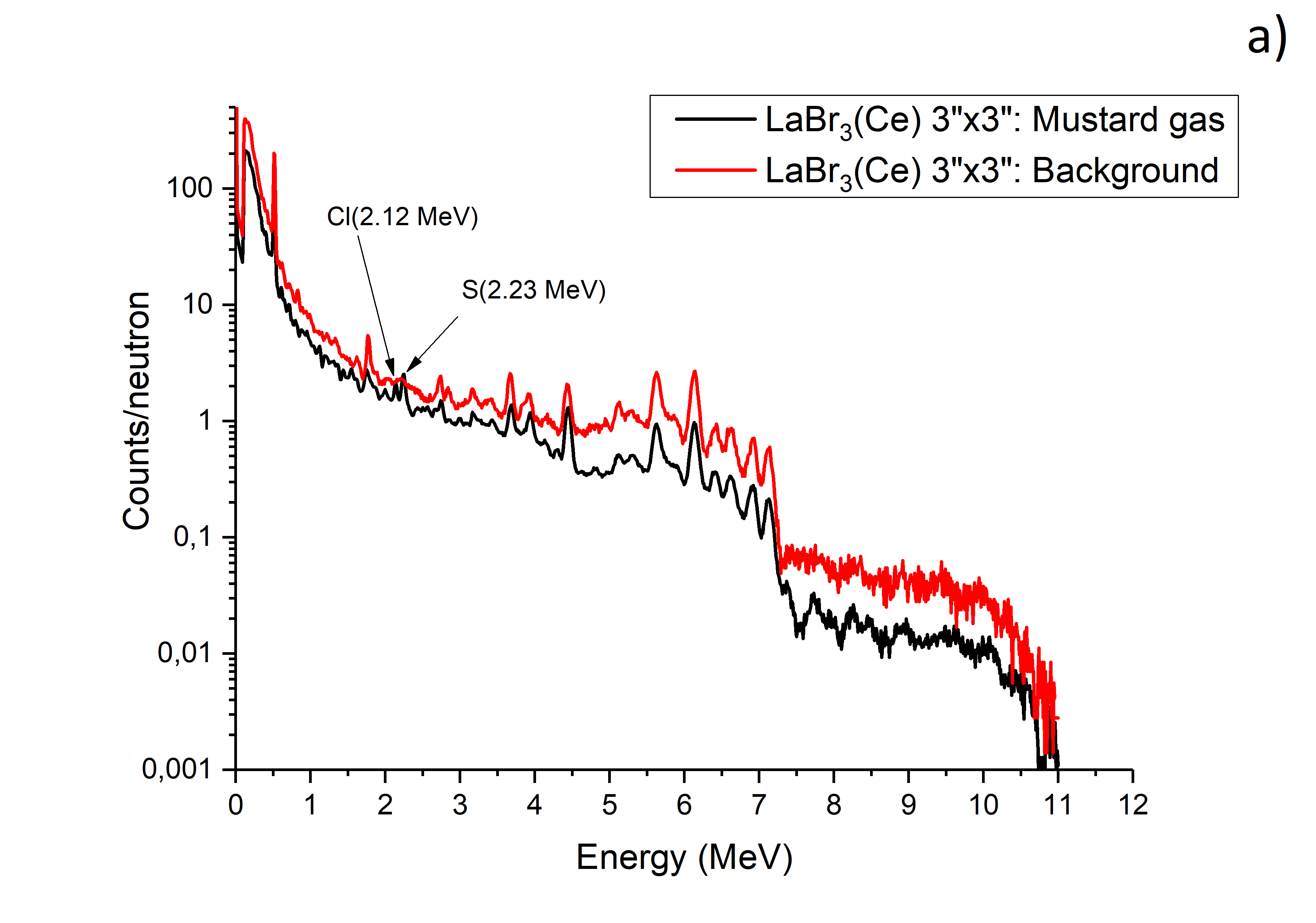}
\includegraphics[width=11.0cm]{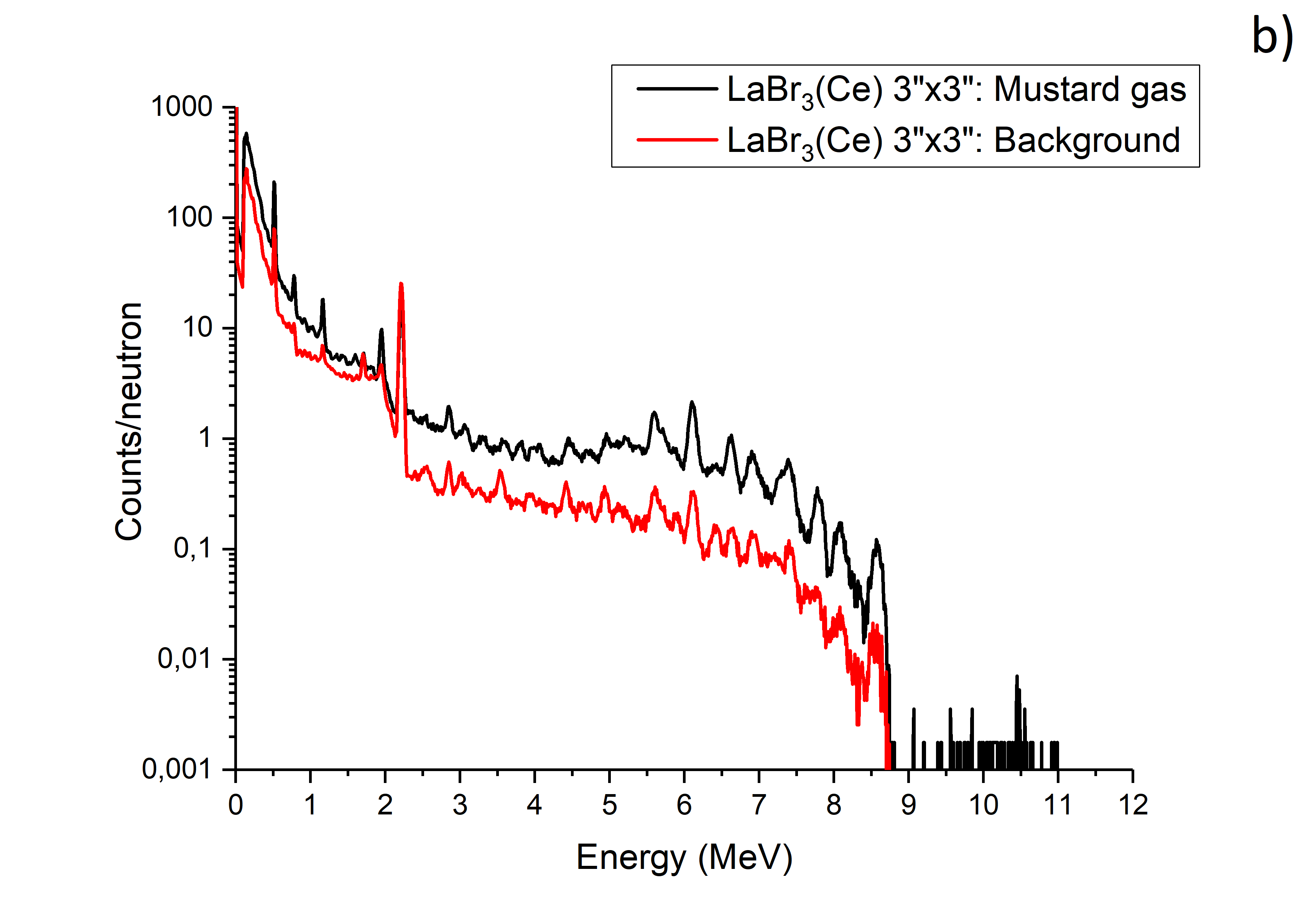}
%}
\caption{Energy distributions simulated assuming the $\gamma$ quantum registration time a) within 50~ns
after neutron emission and b) bigger than 2~$\mu$s after neutron emission. The detector is simulated as
3"x3" LaBr$_3$(Ce) scintillator. The black curve represents simulations with mustard
gas container while in red we present background simulation when the gas is replaced by sand.
}
\label{fig3}
\end{figure}
Fig.~\ref{fig3} a) shows the energy spectra of gamma quanta registered by the detector in the 50~ns
time window after the neutron emission for the mustard gas and background. Surprisingly one can observe
noise reduction in the energy
region of the sulfur peak. Moreover, it turns out that our detection system is even sensitive
to another Cl line in this energy range. The 4.4~MeV line of carbon is unfortunately again merged with Si line.
Chlorine is visible also for delayed $\gamma$ quanta shown in Fig.~\ref{fig3} b) which were registered
2~$\mu$s after neutron emission. This time window provides excellent selectivity of many Cl lines
(1.16, 7.42, 7.80 and 8.58 MeV).
\section{Conclusions and outlook}
We have been developing a new system for underwater threats detection based on the Neutron Activation
Analysis. This device has been optimized in particular to detect chemical agents like mustard gas, soman,
sarin etc. To find the best geometry and materials to build the sensor we have performed Monte Carlo
simulations using the MCNP software. The first preliminary results show that despite large environmental
noise from the water and sand on the bottom of the sea we are able to detect the signature of the hidden
mustard gas. According to the simulations, the characteristic lines of
chlorine are well visible at low energies, while carbon and sulfur are overwhelmed by background.
It turns out that 
%shielding of the $\gamma$ quanta detector from background photons and neutrons does
%not improve sensitivity of the device and
the energy resolution of NaI(Tl) detector is to poor to give
satisfactory results. Thus, the best material to replace semiconductor detector qualified to be LaBr$_3$(Ce)
providing not only good energy and time resolutions, but also reasonable detection efficiency. 
Analysis of the arrival time distribution of measured gamma quanta shows that even rough time gating
gives very promising results. For prompt photons we are able to measure the sulfur and chlorine lines
originating from the mustard gas. Unfortuntely, the 4.5~MeV line of Si overlap with carbon
line around 4.4~MeV. The delayed gamma quanta spectrum shows
excellent selectivity of many Cl lines (1.16, 7.42, 7.80 and 8.58 MeV). Altogether, we are sensitive
to at least two elements of mustard gas which may provide stoichiometry identification.
As it was mentioned before,
the noise originating from the environment can be significantly reduced by the requirement
of the coincident detection of the alpha particles which are produced together with
neutrons~\cite{ActaA}. The $\alpha$ particle detection and its registration time measurement
will be included as the next step in the simulations. This will allow for a final design of the SABAT
sensor.
\section*{Acknowledgments}
This work was supported by the Polish Ministry of Science and Higher Education through
grant No. 7150/E-338/M/2017 and by the National Centre for Research and Development through
grant No. LIDER/17/0046/L-7/15/NCBR/2016.

\end{document}